\documentclass[aps,twocolumn,amsmath,amssymb,showpacs,showkeys,floatfix,a4paper,
nofootinbib]{revtex4}

\usepackage{epsfig}
\usepackage{amssymb}
\usepackage[ansinew]{inputenc}
\usepackage{soul,color}
\usepackage{graphicx}
\usepackage{dcolumn}
\usepackage{bm}

\usepackage{float}
\usepackage{mathrsfs}
\usepackage{tabularx}
\usepackage{midfloat}

\newcommand{\be}{\begin{equation}}
\newcommand{\ee}{\end{equation}}
\newcommand{\bea}{\begin{eqnarray}}
\newcommand{\eea}{\end{eqnarray}}
\newcommand{\ba}{\begin{array}}
\newcommand{\ea}{\end{array}}
\newcommand{\bi}{\begin{itemize}}
\newcommand{\ei}{\end{itemize}}
\newcommand{\lan}{\langle}
\newcommand{\ran}{\rangle}

\renewcommand{\vec}[1]{\mbox{\boldmath $#1 \!\!$ \unboldmath}}

\begin{document}

\title{Sequential Heavy Ion Double Charge Exchange Reactions and the Link to Double $\beta$-decay}

\author{Jessica I.Bellone$^{1}$ {\thanks{Electronic address: jessica.bellone@ct.infn.it}}}
\author{Stefano Burrello$^{1}$ {\thanks{Electronic address:burrello@lns.infn.it}}}
\author{Maria Colonna$^{1}$ {\thanks{Electronic address: colonna@lns.infn.it}}}
\author{Jos\'e-Antonio Lay$^{2}$ {\thanks{Electronic address: lay@us.es}}}
\author{Horst Lenske$^{3}${\thanks{Electronic address: horst.lenske@theo.physik.uni-giessen.de}}}

\affiliation{
$^1$INFN-LNS, I-95123 Catania, Italy
\\
$^2$Departamento de FAMN, Universidad de Sevilla, Apartado 1065, E-41080 Sevilla, Spain\\
$^3$Institut f\"{u}r Theoretische Physik, Justus-Liebig-Universit\"{a}t Giessen, D-35392 Giessen, Germany 
}

\date{\today}

\begin{abstract}
\begin{center}
(NUMEN Collaboration)
\end{center}

Heavy ion double charge exchange reactions are described by sequential meson-exchange, corresponding to a double single charge exchange (DSCE) reaction mechanism. The theoretical formulation is discussed. The fully quantum mechanical distorted wave 2-step calculations are shown to be reproduced very well by approximating the intermediate propagator by its pole part. The role of ion-ion elastic interactions is discussed. As a first application, calculations are performed for the reaction $^{40}$Ca $(^{18}$O $,^{18}$Ne $)^{40}$Ar at 15 AMeV. Results are compared to the data measured at LNS by the NUMEN Collaboration. The common aspects of DSCE reactions and double $\beta$-  decay are discussed by a detailed comparison of the respective
nuclear matrix elements (NME).
\end{abstract}
\pacs {21.60-n,21.60Jz,21.10.Dr}
\keywords{Heavy ion reaction theory, double charge exchange reactions, theory of nuclear charge exchange excitations, {double} beta decay}
\maketitle

\section{Introduction}\label{sec:Intro}
Nuclear double charge exchange (DCE) reactions are of large current interest after it was realized that they give access to a hitherto hardly explored sector of nuclear excitations. In early DCE studies, the focus was on aspects of the dynamics of proton and neutron pair transfer \cite{Dasso:1985zvv,Dasso:1986zza} which at that time was thought to be the dominant reaction mechanism of heavy ion DCE scattering. About a decade later, Blomgren et al. \cite{Blomgren:1995cux} attempted to measure the double-Gamow-Teller resonance (DGTR) in a heavy ion DCE reaction which, however, at that time was not successful. Only recently, it was realized that under appropriate conditions DCE reactions are the perfect tool for spectroscopic nuclear structure investigations \cite{EPJA2018,Takaki:2014}, being also of high interest for the nuclear structure aspects underlying exotic weak interaction processes. That change of paradigm relies on the observation that under appropriate conditions isovector nucleon-nucleon (NN) interactions will be the driving forces, thus extending the longstanding experience with single charge exchange reactions \cite{Len89,Bre88,Boh88,Len98} to higher order processes. By obvious reasons, that conjecture can be explored the best by peripheral coherent reactions with complex nuclei, leading to ejectiles with particle-stable $\Delta Z=\pm 2$ final states. A distinct advantage of heavy ion scattering over the former ($\pi^+,\pi^-$)-DCE reactions \cite{Auerbach:1988ir,Watson:1991ip,Johnson:1994na} is the much easier experimental availability and handling of ion beams.

In this work, we propose a new reaction mechanism for peripheral heavy ion DCE reactions at energies well above the Coulomb barrier. We investigate the conditions under which such reactions can be described as a double single charge exchange (DSCE) process, driven by collisional NN interactions, thus extending our investigations in Refs. \cite{Lenske:2018jav,Lenske:2019rce} to higher order processes. We will not consider
transfer DCE 
which, in fact, has been found to be negligible for the reactions considered 
here \cite{LayBurrello,Burrello:2019jfu}. A formalism is developed for the description of DCE reactions by two consecutive $\Delta Z=\pm 1$ SCE steps. In a DCE reaction, however, the SCE processes are contributing off-the-energy shell as intermediate processes. Hence, their description requires special attention. An important point is the proper treatment of the strongly absorptive elastic ion-ion interactions for which we use a microscopic optical model potential. The spectroscopic aspects are described by Hartree-Fock-Bogoliubov (HFB) and Quasiparticle Random Phase (QRPA) theory following the microscopic approach presented in \cite{Lenske:2018jav}.

As a new aspect, we consider here the connection of DCE reactions and second order weak processes. An almost natural weak counterpart is $2\nu 2\beta$ decay. We will show that
the DCE reaction amplitudes have a striking similarity to the nuclear matrix elements (NME) of $2\nu 2\beta$ decay. While the latter are rare events, heavy ion DCE reactions can be studied frequently under well defined laboratory conditions. In particular, the feasibility of measuring a heavy ion DCE reaction has been recently proved, on the example of the reaction $^{18}$O $+^{40}$Ca $\to ^{18}$Ne $+ ^{40}$Ar at
E/A = 15 AMeV, indeed hinting
towards a direct mechanism \cite{EPJA2015}.  A complication to be dealt with in a DCE reaction is the convolution of the SCE spectra of the intermediate projectile- and target-like nuclei. Since that occurs at half off-shell conditions, the knowledge of the corresponding on-shell SCE cross sections is only of limited advantage. On the theoretical level, the problem is well under control as will be seen by the results discussed below, although inevitably hampered by a certain degree of model dependence -- as is true for SCE reactions with light and heavy projectiles as well, where appropriate methods are available for the extraction of single--beta decay NME \cite{Goodman80,Osterfeld,Taddeucci87,Frekers:2018,Lenske:2018jav}.

The paper is organized as it follows:
In section \ref{sec:DCEforms} the theoretical framework for DSCE reactions is presented. Two descriptions are discussed which emphasize different aspects of sequential DCE reactions. Results of numerical calculations for the afore mentioned reaction $^{18}$O $+^{40}$Ca $\to ^{18}$Ne $+ ^{40}$Ar are discussed in section \ref{sec:Results}. In section \ref{sec:NME} the connection to double $\beta$-decay is established. The paper closes with a summary, conclusions, and an outlook in section \ref{sec:SumOut}.

\section{Theoretical framework of DSCE reactions}\label{sec:DCEforms}

\subsection{General Aspects of Two-Step DCE Reactions}
The formalism developed here
applies to heavy ion DCE reactions of the kind
\be\label{eq:reaction}
^a_za+{}^A_ZA\to{}^a_{z\pm 2}b+{}^A_{Z\mp 2}B
\ee
with special emphasis on the collisional NN-mechanism.

\begin{figure}
\centering\includegraphics[width=\linewidth]{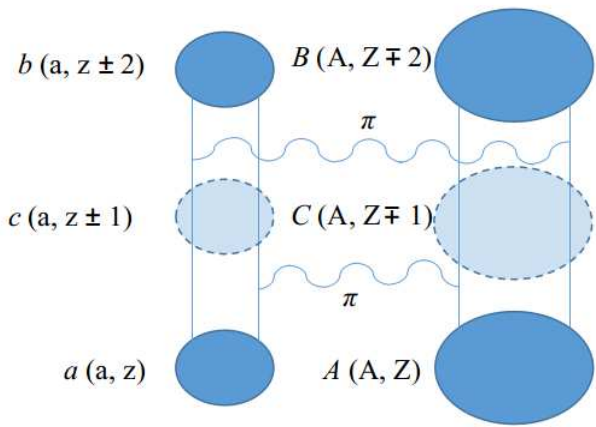}
\caption{(Color online)
Schematic representation of a two-step double charge exchange transition as two consecutive single charge exchange processes.
}
\label{fig:scheme}
\end{figure}

The reaction, leading from the entrance channel $\alpha=\left\{ a,A \right\}$  to the exit channel $\beta=\left\{ b,B \right\}$, changes the charge partition by a balanced redistribution of protons and neutrons. The two reaction partners are acting mutually as the source or sink, respectively, of the charge-transferring virtual meson fields, as depicted in Fig. \ref{fig:scheme}.

The differential DCE cross section
is defined as
\bea\label{eq:xsec_gen}
&&d\sigma_{\alpha\beta}=\frac{m_\alpha m_\beta}{(2\pi\hbar^2)^2}\frac{k_\beta}{k_\alpha}\frac{1}{(2J_a+1)(2J_A+1)}
\nonumber \\
&&\times\sum_{\substack{{\cal M}_a,{\cal M}_A\in \alpha\\{\cal M}_b,{\cal M}_B\in \beta}}{\left|{M^{DCE}_{\alpha\beta}(\mathbf{k}_\alpha,\mathbf{k}_\beta)}\right|^2}d\Omega,
\eea
where  $\mathbf{k}_\alpha$ ($\mathbf{k}_\beta$) denotes the relative 3-momentum and $m_\alpha$ ($m_\beta$) is the reduced mass.
$\{J_a{\cal M}_a,J_A{\cal M}_A\cdots\}$ and $\{J_b{\cal M}_b,J_B{\cal M}_B\cdots\}$ account for the full set of (intrinsic) quantum numbers specifying the initial and final channel
states, respectively.

The DCE reaction mechanism is assumed as a sequence of two uncorrelated SCE events, each one mediated by the action of the isovector NN-interactions, acting between projectile and target and leading to $pn^{-1}$ and $np^{-1}$ particle-hole excitations or vice versa, respectively.  After the first SCE event the system propagates undisturbed until the second interaction. Thus, the reaction proceeds as a double single charge exchange process, which by the number of separate projectile-target interactions is a two-step reaction \cite{Lenske:2018dkz,Lenske:2019var}.

The reaction matrix element, connecting incident and final channels
is readily written down as a quantum mechanical amplitude
in distorted wave approximation (DWA):
\begin{equation}\label{eq:splitSCE}
\begin{split}
&\mathcal{M}^{DSCE}_{\alpha\beta}(\mathbf{k}_{\alpha},\mathbf{k}_{\beta}) {\approx}
\langle\chi^{(-)}_\beta, bB\vert\mathcal{{T}}_{NN}\mathcal{G}
\mathcal{{T}}_{NN}\vert aA,\chi^{(+)}_\alpha \rangle,\\
\end{split}
\end{equation}
corresponding to second order perturbation theory in the residual charge-transferring interaction $T_{NN}$ but being non-perturbative in the initial state (ISI) and final state (FSI) ion-ion interactions. The latter are accounted for, to all orders, by the distorted waves $\chi^{(\pm)}_{\alpha,\beta}({\bf{r}})$ with asymptotically outgoing and incoming spherical waves, respectively.

The anti-symmetrized nucleon-nucleon T-matrix $\mathcal{T}_{NN}$ was discussed in breadth in \cite{Lenske:2018jav}. Central and rank-2 tensor interactions  are included, covering the full spectrum of spin-independent Fermi-type ($S=0,T=1$) and spin-dependent Gamow-Teller-type ($S=1,T=1$) operators of all multipolarities.

The off-shell propagation of the system in the intermediate $\Delta Z=\pm 1$ channels is described by the full many-body Green's function $\mathcal{G}$, given by the eigenstates of the intermediate projectile-like ($c$) and target-like ($C$) nuclei as
\begin{equation}\label{G_gamma}
\mathcal{G}=\sum_{\gamma=cC} \vert cC \rangle G^{(+)}_{\gamma}(\omega_{\alpha})\langle cC \vert .
\end{equation}
The relative motion degrees of freedom are described by the channel Green's functions with asymptotically outgoing spherical waves
\be\label{eq:Gch}
G^{(+)}_{\gamma}(\omega_{\alpha})=\int \frac{d^3k_\gamma}{(2\pi)^3}|\chi^{(+)}_\gamma\rangle
\frac{1}{\omega^{(+)}_\alpha-\omega_\gamma}\langle \tilde{\chi}^{(+)}_\gamma |
\ee
where $\omega^{(+)}_\alpha=\omega_\alpha+i0+$ is located in the upper half of the complex plane, see e.g. \cite{Sat83}.
The energy denominator depends on the total center-of-mass energies of the system in the entrance and intermediate channels, respectively. In non-relativistic notation, we have
\begin{equation}
\omega_{\alpha}=M_a+M_A+\frac{k^2_{\alpha}}{2m_{\alpha}}\quad\quad
\omega_{\gamma}=M_c+M_C+\frac{k^2_{\gamma}}{2m_{\gamma}}.
\end{equation}
where $\omega_\alpha=\sqrt{s_\alpha}$ is fixed by the Mandelstam variable $s_\alpha$.
$M_a$, $M_{A}$ (and $M_c$, $M_C$) denote the nuclear masses in the
initial and intermediate channel, the latter including
excitation energies. $\mathbf{k}_{\gamma}$ indicates the (off-shell) relative momentum in the
intermediate channel.

The Green function is
given by a bi-orthogonal set of distorted waves $\chi^{(\pm)}_\gamma$ and their dual counterparts $\tilde\chi^{(\pm)}_\gamma$ \cite{Sat83}, accounting properly for the elastic ion-ion-interactions with diffractive and strongly absorptive potential components.

We apply to the right hand side of the Eq.\eqref{eq:Gch} the completeness relation
\be\label{eq:CmplRDW}
\int |\tilde\chi^{(-)}_\gamma\ran \frac{d^3k_\gamma}{(2\pi)^3}\lan \chi^{(-)}_\gamma| =1
\ee
and use
\be
\lan \tilde\chi^{(+)}_\gamma|\tilde\chi^{(-)}_\lambda\ran=(2\pi)^3\tilde{S}^\dag_\gamma(\mathbf{k}_\gamma)\delta_{\gamma\lambda}\delta(\mathbf{k}_\gamma-\mathbf{k}_\lambda)
\ee
where $\tilde S_\gamma$ is the dual S-matrix associated with
the hermitian conjugate channel Hamiltonian, i.e. with a creative optical potential. Thus, the channel propagator becomes
\be\label{eq:Gch2}
G_{\gamma}(\omega_{\alpha})=\int \frac{d^3k_\gamma}{(2\pi)^3}|\chi^{(+)}_\gamma\rangle
\frac{\tilde{S}^\dag_\gamma(\mathbf{k}_\gamma)}{\omega^{(+)}_\alpha-\omega_\gamma}\langle {\chi}^{(-)}_\gamma |.
\ee
Inserting Eq.\eqref{eq:Gch2} into Eq.\eqref{eq:splitSCE}, the DSCE transition matrix element reads
$$
\mathcal{M}^{DSCE}_{\alpha\beta}(\mathbf{k}_{\alpha},\mathbf{k}_{\beta})=\sum_{\gamma {=c,C}}\int \frac{d^3k_{\gamma}}{(2\pi)^3}\,
$$
\begin{equation}\label{eq:Mab}
\times \mathcal{M}^{SCE}_{\gamma\beta}(\mathbf{k}_{\gamma},\mathbf{k}_{\beta})
\frac{\tilde{S}^\dag_\gamma(\mathbf{k}_\gamma)}{\omega^{(+)}_\alpha-\omega_\gamma}\mathcal{M} ^{SCE}_{\alpha\gamma}(\mathbf{k}_{\alpha},\mathbf{k}_{\gamma})
\end{equation}
showing that the DCE transition amplitude can be expressed as superposition of reaction amplitudes
$\mathcal{M}^{SCE}_{\alpha\gamma}$ and $\mathcal{M}^{SCE}_{\beta\gamma}$, into and out of the intermediate channels $\gamma$, respectively.

\subsection{The Convolution Approach}

The Cauchy principal value parts of the DSCE amplitudes have the tendency to be suppressed because of compensating positive and negative contributions. To a good approximation, they can be neglected and we may evaluate the convolution integral of the two amplitudes in Pole Approximation (PA), amounting to project the modulus $k_\gamma$  to its on-shell value, defined by $\omega_\gamma=\omega_\alpha$:
\begin{equation}
\begin{split}
&\mathcal{M}^{DSCE}_{\alpha\beta}(\mathbf{k}_{\alpha},\mathbf{k}_{\beta}) \approx -i\pi\sum_{\gamma {= c,C}}k_\gamma m_\gamma\\&
\times \int \frac{d\Omega_\gamma}{(2\pi)^3} \mathcal{M}^{SCE}_{\gamma\beta}(\mathbf{k}_{\gamma},\mathbf{k}_{\beta})
\tilde{S}^\dag_\gamma(\mathbf{k}_{\gamma})
{\mathcal{M}}^{SCE}_{\alpha\gamma}(\mathbf{k}_{\alpha},\mathbf{k}_{\gamma}).
\end{split}
\label{eq:MDCE_CA}
\end{equation}
This kind of approach maintains the character of the DCE reaction as a sequence of two independent SCE reactions. In PA the reaction amplitude displays that property by the convolution of two on-shell SCE amplitudes which, in principle, are accessible in SCE reactions. However, in practice this would mean to identify SCE transition up to high excitation energies of close to 100 MeV.

As seen below, Eq.\eqref{eq:MDCE_CA}, leads to an astonishingly good reproduction of the full two-step DW cross sections. However, for the sake of a deeper insight into the essentials of the DSCE reaction mechanism, further reductions are extremely valuable. For example, additional steps are necessary for the extraction of spectroscopic information out of measured cross sections, because from Eq.\eqref{eq:MDCE_CA} 
the relation of the DSCE reaction amplitude to projectile and target
nuclear matrix elements is not immediately clear. 
A caveat is the presence of initial state and final state interactions.
We also note that although the intermediate SCE amplitudes appear to be of DWA-type, they are in fact half--off--shell quantities which as such cannot be measured independently.

In order to quantify those effects a separation of elastic ion-ion interactions and nuclear structure effects is helpful.
In momentum representation, the SCE amplitudes are given as \cite{Lenske:2018jav}
\be\label{eq:MSCE}
\mathcal{M}^{SCE}_{\alpha\gamma}=\int d^3p \mathcal{N}_{\alpha\gamma}(\mathbf{p},\mathbf{k}_\alpha,\mathbf{k}_\gamma)\mathcal{U}^{SCE}_{\alpha\gamma}(\mathbf{p})
\ee
and for the second SCE amplitude accordingly.
The transition potential $\mathcal{U}^{SCE}_{\alpha\gamma}=\lan \varphi_{\mathbf{k}'},cC|\mathcal{T}_{NN}|aA,\varphi_{\mathbf{k}}\ran$ corresponds to the reaction amplitudes evaluated with plane waves $\varphi_{\mathbf{k}}$ and $\mathbf{p}=\mathbf{k}-\mathbf{k}'$. Their structure for central interactions is
\bea\label{eq:Uab}
&&\mathcal{U}^{SCE}_{\alpha\gamma}(\mathbf{p})=\sum_{S=0,1,T=1}V^{(C)}_{ST}(p^2) \nonumber \\
&&\times \lan c| \mathcal{R}_{ST}(\mathbf{p},1_a)|a\ran\cdot
\lan C|\mathcal{\mathcal{R}}_{ST}(\mathbf{p},2_A)|A\ran,
\eea
where the bilinear forms of one-body operators \cite{Lenske:2018jav}
\be\label{eq:RST}
\mathcal{R}_{ST}(\mathbf{p},k)=e^{i\mathbf{p}\cdot \mathbf{r}_k}\left(\bm{\sigma}_k\right)^S\left(\bm{\tau}_k\right)^T ,
\ee
acting in projectile ($k$=1) or target ($k$ = 2), respectively, have been introduced.
Expressions for rank-2 spin-tensor interactions are found in \cite{Lenske:2018jav}.

Elastic ion-ion interactions are accounted for by the distortion coefficient
\be
\mathcal{N}_{\alpha\gamma}(\mathbf{p},\mathbf{k}_\alpha,\mathbf{k}_\gamma)=\frac{1}{(2\pi)^3}
\lan \chi^{(-)}_\gamma |e^{i\mathbf{p}\cdot \mathbf{r}}|\chi^{(+)}_\alpha\ran \nonumber
\ee
which can be considered as an off-shell extension of the S-matrix, approaching in the plane wave (PW) limit  $\mathcal{N}^{(PW)}_{\alpha\gamma}=\delta(\mathbf{p}+\mathbf{k}_\alpha-\mathbf{k}_\gamma)$.
In \cite{Lenske:2018jav}, the distortion coefficients were investigated in detail for SCE reactions. For the present case, it is important that $\mathcal{N}_{\alpha\gamma}$ can be decomposed into a forward component, given by an absorption factor $n_{\alpha\gamma}$ and a residual distortion form factor which we neglect in the following. Hence, we use in forward scattering approximation
\be
\mathcal{N}_{\alpha\gamma}(\mathbf{k}_\alpha,\mathbf{k}_\gamma)\simeq n_{\alpha\gamma}\delta(\mathbf{p}+\mathbf{k}_\alpha-\mathbf{k}_\gamma).
\ee
That leads in Eq.\eqref{eq:Mab}
to a product of two distortion residues and the dual S-matrix. The absorptive effects from the intermediate channels are cancelled to a large extent by the dual S-matrix which allows to replace the product of the two distortion coefficients and the dual S-matrix by the residue $N_{\alpha\beta}=\lan n_{\gamma \beta}\tilde{S}^\dag_\gamma n_{\alpha\gamma}\ran_\gamma$, appropriately averaged over the intermediate channels.
Thus, at low momentum transfer the DSCE amplitude is given approximately by
\begin{equation}
\begin{split}
&\mathcal{M}^{DSCE}_{\alpha\beta}(\mathbf{k}_{\alpha},\mathbf{k}_{\beta}) \approx N_{\alpha\beta}(\mathbf{k}_{\alpha},\mathbf{k}_{\beta})\sum_{\gamma {= c,C}}\\&
\times \int \frac{d^3k_\gamma}{(2\pi)^3} \mathcal{U}^{SCE}_{\gamma\beta}(\mathbf{k}_{\gamma}-\mathbf{k}_{\beta})\frac{1}{\omega_\alpha-\omega_\gamma+i\eta}
{\mathcal{U}}^{SCE}_{\alpha\gamma}(\mathbf{k}_{\alpha}-\mathbf{k}_{\gamma})
\end{split}
\label{eq:UDSCE}
\end{equation}
by which one can separate 
nuclear and reaction dynamics. The expression can be reduced further by the pole approximation as in Eq.\eqref{eq:MDCE_CA}. Finally, at small momentum
transfer,
ISI and FSI effects are restored into the matrix elements by replacing the PW amplitudes $\mathcal{U}^{SCE}_{\kappa\lambda}$ by amplitudes evaluated with distorted waves in the incoming $\alpha$ and the outgoing $\beta$ channel, but retaining the plane waves in the intermediate channels $\gamma$.

\subsection{The Separation Approach}
In this section, an approach is presented which allows the separation of the DCE reaction amplitude into a nuclear structure and a reaction part by exploiting the distorted wave completeness relation. That requires to go somewhat deeper into the multipole structure of DCE reaction amplitudes. The SCE transition form factors,
expressed as a function of the distance between projectile
and target centers of mass, read:
\be
F_{\alpha\gamma}(\mathbf{r})=\sum_{S=0,1}\lan cC|V^{(C)}_{S_\lambda T}(\vec{\sigma}_a\cdot \vec{\sigma}_A)^{S}\vec{\tau}_a\cdot \vec{\tau}_A|aA\ran
\ee
They are expanded into multipole form factors
\be
F_{\alpha\gamma}(\mathbf{r})\sim
\sum_{\lambda\mu}\sum_{\lambda_c\lambda_C}A^{\lambda_c\lambda_C}_{S\lambda}F_{(S\lambda_c\lambda_C)\lambda\mu}(\mathbf{r})
\ee
where angular momentum coupling coefficients involving the nuclear spins are not shown explicitly, but for which we refer to Ref. \cite{Lenske:2018jav}.
The A-coefficients contain the remaining coupling of the projectile and target multipoles to the resulting total angular momentum transfer $\lambda$ with projection $\mu$. The form factors
are parameterized in terms of transition amplitudes $\beta_{S\lambda_k}$
and reduced form factors of unit transition strength:
\be
F_{(S\lambda_c\lambda_C)\lambda\mu}(\mathbf{r})=\left[\beta^{ac}_{S\lambda_c} \beta^{AC}_{S\lambda_C} \right]U_{S\lambda\mu}(\mathbf{r}).
\ee
as practiced successfully in the
Multi Step Direct Reaction (MSDR)-theory of \cite{Tamura:1982zzv,Lenske:1983tnq,Lenske:2001rfn}. For practical purposes, it is useful to define the coupled spectroscopic amplitudes
\be
\beta^{ac,AC}_{S\lambda}=\sum_{\lambda_c\lambda_C}A^{\lambda_c\lambda_C}_{S\lambda}\beta^{ac}_{S\lambda_c} \beta^{AC}_{S\lambda_C}
\ee
With corresponding expressions for the second step form factor, the summation over the intermediate states leads to the spectroscopic densities
\bea
&&\rho^{S_{1}S_{2}}_{\lambda_1\lambda_2}(\omega_\alpha,k_\gamma) \nonumber \\
&&=\sum_{cC} \frac{\beta^{cb,CB}_{S_{2}\lambda_2}\beta^{ac,AC}_{S_{1}\lambda_1}}{\omega^{(+)}_\alpha-M_c-M_C-k^2_\gamma/2m_\gamma}
\eea
In the energy denominator, we replace $k_\gamma$ by an average value $\bar k$. By this manipulation, the propagator and the $k_\gamma$--integration are decoupled. The latter leads to the DW completeness
relation in the intermediate channel.
The reaction amplitude becomes
\bea\label{eq:MDCE_SA}
\mathcal{M}^{DSCE}_{\alpha\beta}&&\approx \sum_{S_{1}\lambda_1\mu_1,S_{2}\lambda_2\mu_2}\rho^{S_{1}S_{2}}_{\lambda_1,\lambda_2}(\omega_\alpha,\bar k)\nonumber\\
&&\times \lan \chi^{(-)}_\beta|U_{S_2\lambda_2\mu_2}U_{S_1\lambda_1\mu_1}|\chi^{(+)}_\alpha\ran .
\eea
An irreducible representation is obtained by further coupling the two reduced form factors to total angular momenta. Thus, the DCE process is described by a reduced DW reaction amplitude fully accounting for the reaction dynamics. The two-step character of the DCE process leads to the special kind of form factors. The nuclear
structure aspects of the sequential DCE process are contained in the spectroscopic density combining the SCE response of projectile and target. The $\beta$-amplitudes are related to the reduced beta-decay matrix elements in the same way as known from the so-called collective model for inelastic scattering \cite{Sat83}. Theoretically, they are fixed by the nuclear SCE response functions as fractions of multipole sum rules which in parallel determines also the reduced form factors $U_{S\lambda\mu}$ \cite{Tamura:1982zzv,Lenske:1983tnq,Lenske:2001rfn}.

\section{Results}\label{sec:Results}
\subsection{Numerical details}
The theory is of general applicability without constraints neither on the kind of transition nor on the multipolarity.
In the reaction calculations, microscopic optical potentials are used. They were obtained by folding the Hartree-Fock-Bogolubov (HFB) one-body ground-state densities of projectile and target with the isoscalar and isovector parts of the (anti-symmetrized) NN T-matrix of Ref.\cite{FraneyLove}.
As discussed in \cite{Lenske:2018jav,Lenske:2019rce} QRPA calculations are performed to evaluate
the SCE projectile and target transition densities. The projectile and target transition form factors were obtained by folding the QRPA transition densities with the central and rank-2 tensor parts of the anti-symmetrized NN T-matrix of Ref. \cite{FraneyLove}. All folding calculations were done in momentum representation.

Two kinds of reaction calculations were performed: The full partial wave two-step formalism, as discussed e.g. in \cite{Tamura:1982zzv,Lenske:1983tnq,Lenske:2001rfn}, was used in solving the set of inhomogeneous scattering equations by direct numerical integration, as available by the computer code
FRESCO \cite{Ian}.
These results serve as benchmark calculations for the approximations discussed in section \ref{sec:DCEforms}. In parallel, independent calculations using the perturbation theoretical formalism were performed: Single charge exchange form factors, the DWA reaction amplitudes, and the corresponding cross sections
were calculated by our standard DWA-SCE computer code package
HIDEX \cite{Cappuzzello:2004afa}.
As found in Ref.\cite{Lenske:2018jav}, the calculation of the SCE reaction amplitudes requires a quite involved angular momentum algebra 
to couple the intrinsic nuclear angular momenta to the resulting total orbital angular momentum, by which the multipolarity observed at the level of the cross section is determined.
In general, one finds
\bea \label{eq:Irrep}
&&\mathcal{M}^{SCE}_{\alpha\beta }(\mathbf{k}_\alpha,\mathbf{k}_\beta)=\sum_{\ell_\alpha,\ell_\beta;\ell m}
C^{\ell_\alpha\ell_\beta \ell}_{J_aJ_AJ_bJ_B}\nonumber \\
&&\times M_{\ell_\alpha\ell_\beta\ell}(k_\alpha,k_\beta)
\left[Y_{\ell_\alpha}(\Omega_\alpha)Y_{\ell_\beta}(\Omega_\beta)\right]_{\ell m}
\eea
where the C-coefficients describe the recoupling of nuclear spins $J_{a,b}$ and $J_{A,B}$, respectively, to the angular momenta $\ell$ 
acting in the ion-ion relative motion sector.
The situation simplifies, however, for the $(0^+_a,0^+_A)\to (0^+_b,0^+_B)$ case as here. Then, the transitions into and out of the intermediate states necessarily must proceed through the same kind of multipolarity $\ell$,
leading to a total angular momentum transfer $L = 0$.

\begin{figure}
\centering\includegraphics[width=\linewidth]{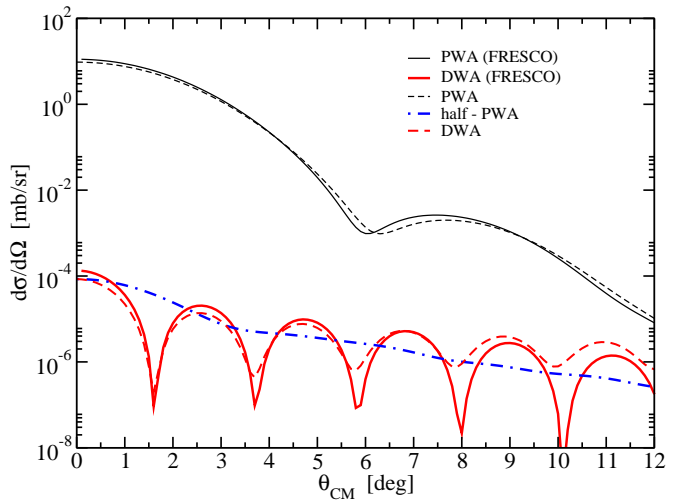}
\caption{(Color online)
Angular distribution of the differential cross section for the DCE reaction $^{18}$O $+ ^{40}$Ca $\to ^{18}$Ne $+ ^{40}$Ar at 15 AMeV, as obtained within PWA
(black thin lines) and DWA (red thick lines).
Only one intermediate channel is considered (see text).
Our simulations (dashed lines) are compared to the results of the FRESCO
code (full lines).
The dot-dashed blue curve refers to an hybrid calculation with plane waves in the intermediate channel, thus modelling the separation approach, Eq.\eqref{eq:UDSCE}.}
\label{fig:DCEx_vs_FRESCO}
\end{figure}

\subsection{The reaction $^{18}O+{}^{40}Ca\to {}^{18}Ne+{}^{40}Ar $}


As a first illustrative application,
we consider the  $^{18}$O $+ ^{40}$Ca $\to ^{18}$Ne $+ ^{40}$Ar reaction at 15 AMeV, which has been the object of recent experimental investigations \cite{EPJA2015}.
We focus on the simplest case, namely $0^+_{gs} \to 0^+_{gs}$ transitions both in projectile and target.
The intermediate channels of this DCE reaction are defined by the odd-odd nuclei $^{18}F$ and $^{40}K$, which are both of a quite complex spectroscopic structure: Rather dense spectra with a number of high-spin levels are observed close
to 
$^{18}F(1^+,g.s.)$ as well as in close vicinity to $^{40}K(4^-,g.s.)$.  A detailed survey of the corresponding spectra and their rather successful description by our charged-current QRPA (ccQRPA) is found in \cite{Lenske:2018jav,Lenske:2019rce}. Here, the QRPA spectral distributions and transition densities for $^{18}O(0^+,g.s.)\to{}^{18}F(J^\pi,E_x)$ and
$^{40}Ca(0^+,g.s.)\to{}^{40}K(J^\pi,E_x)$ are used to calculate form factors, transition potentials, and the SCE reaction amplitudes.

We start following the convolution approach.
For the sake of
simplicity, to check the quality of our calculations and to investigate
the relevance of the distortion effects, we first include
only one single intermediate channel.
In particular, we consider the transition with total angular momentum and parity transfer $J^{\pi}= 1^+$ for both projectile and target nuclei, leading to the ground state of $^{18}F$ and to the first excited $1^+$ state at $E_x=2.29$~MeV of
$^{40}K$.
These states correspond to the Gamow-Teller transitions discussed in
Ref.\cite{Lenske:2018jav}.
The SCE amplitudes were used to construct the second order integral of Eq.\eqref{eq:MDCE_CA} which was evaluated numerically.
An instructive exercise is to compare results of calculations in plane wave approximation (PWA) and in DWA, giving insight on the effects of elastic ion-ion interactions in DCE reactions. In Fig. \ref{fig:DCEx_vs_FRESCO} we show the angular distributions obtained by second order PWA and DWA calculations for the reaction considered.
That figure contains a number of important messages for future research on heavy ion DCE reactions. First of all, ISI and FSI effects suppress cross sections by many orders of magnitudes as a result of the strong absorption, showing that $N_{\alpha\beta}\sim 10^{-5}$.
As discussed in \cite{Lenske:2018jav},
the quenching will increase rapidly with increasing target and/or projectile mass. The suppression decreases with incident energy which, however, at realistically accessible energy scales hardly compensates the mass-dependent quenching.
One can also observe that our numerical calculations,
based on Eq.\eqref{eq:MDCE_CA}, i.e. adopting the PA (dashed lines), reproduce quite well
the FRESCO results (full lines).

The results show a strong influence of the optical model potentials on the diffraction structure of angular distributions, superimposing those reflecting the reaction form factor properties.
In Fig. \ref{fig:DCEx_vs_FRESCO}, this is realized by comparing the DWA results (dashed red line) to
the results which were obtained considering plane
waves in the intermediate channel (dot-dashed blue line). The latter calculation is able to
reproduce the DWA cross section at very small angles, but exhibits a quite
flat angular distribution. These results indicate both the virtue and the limitations of the scaling approach, Eq.\eqref{eq:UDSCE}: At vanishing momentum transfer, the magnitude of the full DWA two-step cross section is rather well described, but the scaling approach
is unable to account for the diffraction structure at larger momentum transfer, {thus restricting that kind of approach to extreme forward angles. }

\begin{figure}
\centering\includegraphics[width=1.\linewidth]{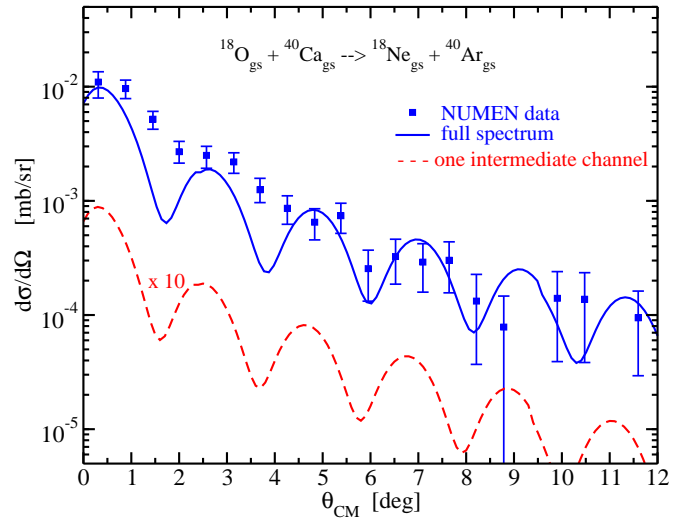}
\caption{(Color online) Experimental angular distribution for the DCE reaction $^{18}$O$_{gs} + ^{40}$Ca$_{gs} \to ^{18}$Ne$_{gs} + ^{40}$Ar$_{gs}$ at 15 AMeV \cite{EPJA2015} compared with DSCE calculation performed
with only one intermediate state (red dashed line),
and considering the full virtual intermediate state integration
(blue full line).
Both cross sections are folded with the experimental angular resolution ($\Delta\theta_{exp}=0.6^\circ$). }
\label{fig:DSCE_vs_data}
\end{figure}

We now move to discuss the results obtained considering an extended spectrum of intermediate states.  We have taken into account intermediate transitions up to
${E_x}$ = 15 MeV and $0^\pm \leq J^\pi \leq 5^\pm $,
for both  $^{18}F$ and $^{40}K$, see \cite{Lenske:2018jav} for more details.
It has been checked that this choice leads
to convergent results.
Calculations have been performed following the (more easy to handle) formalism
provided by Eq.\eqref{eq:MDCE_SA}, where energy conservation has been imposed
to determine the average value ${\bar k}$.
Results are shown in Fig.\ref{fig:DSCE_vs_data} and compared
to the NUMEN experimental {data, where the theoretical angular distributions were folded with the experimental angle resolution.}
The DWA results of Fig.\ref{fig:DCEx_vs_FRESCO}, with only one intermediate state, are also shown
for comparison.
{By the inclusion of the full spectrum of
intermediate channels, the calculations come close to
the data without additional adjustment. However, the diffraction structure of the DWA angular distributions is more pronounced than observed experimentally, which is the cause for the slight apparent underestimation of the data at certain forward angles.
With all care, these remaining discrepancies may indicate contributions of processes of an origin different from the DSCE reaction mechanism, e.g. the two-nucleon scenario discussed in \cite{Lenske:2019rce}.}

\section{Nuclear Matrix Elements and Relation to Double--$\beta$--decay}\label{sec:NME}

On the formal level, the DSCE amplitudes have much in common with the NME of $2\nu 2\beta$ decay because both are of the same kind of second order perturbation theory. There are no strong reasons that the Single State Dominance (SSD) assumption, found to work well for $2\nu 2 \beta$ decay, must hold also for heavy ion DSCE reactions. Quite contrary, despite of the fundamental differences in the dynamics, sequential DCE reactions are surprisingly similar to what is expected for $0\nu 2\beta$ decay. For example, the measured angular range in Fig.\ref{fig:DSCE_vs_data} covers in fact momentum transfers up to $q_{\alpha\beta}\sim 400$MeV/c, implying that the intrinsic system must support these large
momenta. This resembles
the conditions found in theoretical studies of $0\nu 2\beta$ decay, see e.g. \cite{Tomoda:1990rs}. In this sense, heavy ion DCE reactions are of a hybrid character, covering aspects of both kinds of double-beta decay.

Irrespective of those ambiguities, the extraction of nuclear matrix elements off the cross sections is a key task of DCE spectroscopy. For SCE reactions the NME--extraction directly from data relies on the definition of a so-called \emph{unit cross section} meant to include  at the given incident energy besides kinematical factors the SCE interaction strengths and distortion factors \cite{Taddeucci87,Lenske:2018jav}. While that concept is comparatively easy realized for both light and heavy ion SCE reactions, some efforts are required for a DCE reactions. Here, we elucidate the unit cross section problem for two extreme scenarios, namely the SSD assumption and the case of a spectral distribution without a dominant state.

Let us assume that the target-like SCE spectra $A\to C$ and $C\to B$ are dominated by a transition into and out of a single state $R$. Then, a decent approximation is to restrict the summation over the channel states $C$ to the contribution of the dominating target resonance with $M_C=M_R$ but include the full spectrum of states $c$. According to Eq.\eqref{eq:Uab} 
and Eq.\eqref{eq:UDSCE} the latter results in the projectile isovector polarization tensor
\bea
&&\Pi^{S'S}_{ab}(\mathbf{q}_{\gamma\beta},\mathbf{q}_{\alpha\gamma})=\nonumber \\
&&\sum_c\frac{\lan b| \mathcal{R}_{S'T}(\mathbf{q}_{\gamma\beta},1'_a)|c\ran
\lan c| \mathcal{R}_{ST}(\mathbf{q}_{\alpha\gamma},1_a)|a\ran}{\omega^{(+)}_\alpha-M_c-M_R-k^2_\gamma/2m_\gamma}
\eea
Replacing $M_c$ by an average (spectral) mass $\bar{M}_c$ allows to evaluate the summation by closure:
\bea
&&\sum_c\lan b| \mathcal{R}_{S'T}(\mathbf{q}_{\gamma\beta},1'_a)|c\ran
\lan c| \mathcal{R}_{ST}(\mathbf{q}_{\alpha\gamma},1_a)|a\ran\nonumber \\
&&=\lan b|e^{i\mathbf{q}_{\alpha\beta}\cdot \mathbf{r}_a}
(\bm{\sigma}')^{S'}(\bm{\sigma})^{S}\tau'_\pm\tau_\pm|a\ran
\eea
with $\mathbf{r}_a=(\mathbf{r}_1+\mathbf{r}_{1'})/2$. The $a\to b$ matrix elements have to be contracted with the corresponding $A\to R\to B$ matrix elements to a total spin--scalar reaction amplitude:
\bea \label{eq:MDSCE_R}
&&\mathcal{M}^{DSCE}_{\alpha\beta}(\mathbf{k}_\alpha,\mathbf{k}_\beta)\approx N_{\alpha\beta}(\mathbf{k}_\alpha,\mathbf{k}_\beta)\nonumber \\
&&\times\sum_{SS'}V^{S'S}_{\alpha\beta}(\mathbf{k}_\alpha,\mathbf{k}_\beta)
 \lan b|e^{i\mathbf{q}_{\alpha\beta}\cdot \mathbf{r}}
 (\bm{\sigma}')^{S'}(\bm{\sigma})^{S}\tau'_\pm\tau_\pm|a\ran\nonumber \\
&& \otimes
\overline{\lan B|\mathcal{R}_{S'T}(\mathbf{q}_{\gamma_{\bar{c}R}\beta})|R\ran\lan R|\mathcal{R}_{ST}(\mathbf{q}_{\alpha \gamma_{\bar{c}R}})|A\ran}
\eea
In accordance with the assumed dominance of the state $R$, the transition operators are evaluated at momentum transfers involving the effective relative momentum
\be\label{eq:kR}
k^2_{\gamma\bar{c}R}=\frac{1}{4s_\alpha}\left(s_\alpha-(M_R+\bar{M}_c)^2 \right)\left(s_\alpha-(M_R-\bar{M}_c)^2 \right),
\ee
{and the averaging over its direction is indicated in Eq.\eqref{eq:MDSCE_R} by the overline. }

We have introduced the effective DCE interaction given as a matrix in spin-space:
\be\label{eq:VSS}
V^{S'S}_{\alpha\beta}(\mathbf{k}_\alpha,\mathbf{k}_\beta)= \int \frac{d^3k_\gamma}{(2\pi)^3}
\frac{V^{(C)}_{S'T}(\mathbf{q}_{\gamma\beta})V^{(C)}_{ST}(\mathbf{q}_{\alpha\gamma})}{\omega^{(+)}_\alpha-\bar{M}_c-M_R-k^2_\gamma/2m_\gamma},
\ee
which in fact is well described in pole approximation.

The dyadic products of isospin operators are the $I_3=\pm 2$ components of a rank-2 isotensor operator
$\mathbf{I}_2=\left[\bm{\tau}'\otimes \bm{\tau}  \right]_2$. Accordingly, the products of spin-operators are the elements of spin-tensors of rank
$|S-S'|\leq r_S \leq S+S'$.
Simplifications occur for $0^+_A\to 0^+_B$ reactions. If $R$ is a Gamow-Teller or a spin-dipole resonance, then only $S=S'=1$ is allowed.  $S=S'=0$ is selected if $R$ is the Fermi resonance. 
The $S=S'=1$ selection rule applies in particular if $B$ is the DGTR.

If the SSD scenario applies to both nuclei, the reaction amplitude is
\bea \label{eq:MDSCE_rR}
&&\mathcal{M}^{DSCE}_{\alpha\beta}(\mathbf{k}_\alpha,\mathbf{k}_\beta)\approx N_{\alpha\beta}\sum_{SS'}V^{S'S}_{\alpha\beta}(\mathbf{k}_\alpha,\mathbf{k}_\beta)\nonumber \\
&&\times \overline{\lan b|\mathcal{R}_{S'T}(\mathbf{q}_{\gamma_{rR}\beta},1'_a)|r\ran\lan r|\mathcal{R}_{ST}(\mathbf{q}_{\alpha \gamma_{rR}},1_A)|a\ran} \nonumber \\
&&\otimes \overline{\lan B|\mathcal{R}_{S'T}(\mathbf{q}_{\gamma_{rR}\beta},2'_A)|R\ran\lan R|\mathcal{R}_{ST}(\mathbf{q}_{\alpha \gamma_{rR}},2_A)|A\ran}
\eea
and in Eq.\eqref{eq:VSS} one has to replace $\bar{M}_c$ by $M_r$. The matrix elements involve momentum transfers with the effective momentum $k_{\gamma_{rR}}$ which is defined according to Eq.\eqref{eq:kR} but replacing $\bar{M}_c$ by $M_r$.

Lastly, we consider the case that the SSD does not apply to any of the ions. By replacing $M_{c,C}$ by average spectral masses $\bar{M}_{c,C}$ the summations over the intermediate channel states can be evaluated by closure. The resulting reaction amplitude is
\bea \label{eq:MDSCE_CC}
&&\mathcal{M}^{DSCE}_{\alpha\beta}\approx N_{\alpha\beta}\sum_{SS'}V^{S'S}_{\alpha\beta}\nonumber \\
&&\times \lan b|e^{i\mathbf{q}_{\alpha\beta}\cdot \mathbf{r}_a}(\bm{\sigma}')^{S'}(\bm{\sigma})^{S}\tau'_\pm\tau_\pm|a\ran  \nonumber \\
&&\otimes \lan B|e^{i\mathbf{q}_{\alpha\beta}\cdot \mathbf{r}_A}(\bm{\sigma}')^{S'}(\bm{\sigma})^{S}\tau'_\mp\tau_\mp|A\ran
\eea
and Eq.\eqref{eq:VSS} is evaluated with $\bar{M}_C$ instead of $M_R$.

These examples indicate the following interesting leading order scheme:
\begin{itemize}
  \item If SSD is valid in one of the nuclei, in that nucleus a product of two SCE-NME of one-body operators into and out of the dominating state $R$ is obtained, together with a DCE-matrix element describing the transition into the final $|\Delta Z|=2$ configuration of the other nucleus directly by an effective rank-2 isotensor two-body operator -- see Eq.\eqref{eq:MDSCE_R}.
  \item If SSD applies to both nuclei, for each of the nuclei products of two one-body SCE-NME into and out of the states $r$ and $R$ are found  -- see  Eq.\eqref{eq:MDSCE_rR},
  \item If SSD does not apply in both nuclei, the transitions in both nuclei are given by matrix elements of effective isotensor two-body operators -- see Eq.\eqref{eq:MDSCE_CC}.
\end{itemize}
In all cases, the NME's are components of spin-spin tensors which are contracted with the matrix of second order DCE interaction form factors $V^{SS'}_{\alpha\beta}$. According to Eq.\eqref{eq:VSS}, the matrix $V^{SS'}_{\alpha\beta}$ depends on the kinematics and Q-values of the reaction. The elements of that matrix are understood as effective coupling constants for the various non-spinflip and spinflip transitions contributing to the DCE excitations in projectile and target.

An interesting observation is made for those cases where SSD does not apply: For $S=S'=1$ the $|\Delta Z|=2$ NME's are determined by a double GT-operator (DGTO) $\sim \vec{\sigma}\vec{\tau}\vec{\sigma}'\vec{\tau}'$ which can be decomposed into spin-operators of multipolarity $\lambda=0,2$ accompanied by a rank-2 isotensor operator. Hence, the DGTO will contribute to the excitation of the DGTR and the Double Isobaric Analog State (DIAS), and in connection with a p-wave interaction, also to excitations of double spin-dipole modes. The DGT-type of operator is under active scrutiny in sum rule studies of double excitations of charge exchange modes \cite{Roca-Maza:2019gtz,Auerbach:2018byu,Sagawa:2016www,Zheng:1990vg,Vogel:1988ve}. Correspondingly, the double-Fermi operator (DFO) encountered for $S=S'=0$ is of interest for investigations of the DIAS mode.

\subsection{Nuclear Matrix Elements and Unit DCE cross sections}
{Last but not least, we indicate briefly the connection of DCE cross sections to nuclear matrix elements. }
As an important common feature of the reaction amplitudes, the separation into the matrix of second order interactions and nuclear matrix elements was derived, resulting in the structure:
\be
\mathcal{M}^{DSCE}_{\alpha\beta}\approx N_{\alpha\beta}\sum_{SS'}V^{SS'}_{\alpha\beta} F^{SS'}_{ab}F^{SS'}_{AB}
\ee
where $F^{SS'}_{ab}$ and $F^{SS'}_{AB}$ denote the transition form factors according to the discussed cases. At vanishing momentum transfer, these form factors reduce to NME of the $\Delta Z=\pm 2$ excitations in the projectile and target nucleus, respectively. {Their interpretation, however, requires further analysis as emphasized by the discussions in the previous sections. }

Inclusive cross sections like those measured in the pioneering $^{18}O+{}^{40}Ca\to {}^{18}Ne+{}^{40}Ar $ experiment are determined in leading order -- up to interference terms -- by the trace over the spin-spin tensor structures:
\bea
&&d\sigma_{\alpha\beta}\sim |\mathcal{M}^{DSCE}_{\alpha\beta}|^2\sim |N_{\alpha\beta}|^2
\sum_{S_1S_2}\nonumber \\
&&|V^{S_1S_2}_{\alpha\beta}|^2|F^{S_1S_2}_{ab}(\mathbf{q}_{\alpha\beta})|^2|F^{S_1S_2}_{AB}(\mathbf{q}_{\alpha\beta})|^2d\Omega_{\alpha\beta}+...
\eea
Thus, we may introduce the \textit{DCE unit cross sections}
\be
d\bar{\sigma}^{(S_1S_2)}_{\alpha\beta}=\frac{m_\alpha m_\beta}{(2\pi\hbar^2)^2}\frac{k_\beta}{k_\alpha} |N_{\alpha\beta}|^2|V^{S_1S_2}_{\alpha\beta}|^2d\Omega_{\alpha\beta}
\ee
such that the DCE cross section is given as
\bea
&&d\sigma_{\alpha\beta}(\mathbf{q}_{\alpha\beta})\simeq \sum_{SS'}d\bar{\sigma}^{(S_1S_2)}_{\alpha\beta}(\mathbf{q}_{\alpha\beta})\nonumber \\
&&|F^{S_1S_2}_{ab}(\mathbf{q}_{\alpha\beta})|^2|F^{S_1S_2}_{AB}(\mathbf{q}_{\alpha\beta})|^2
d\Omega_{\alpha\beta}
\eea
by which the relations known for SCE cross sections are generalized to DCE reactions.

\section{Summary and Outlook}\label{sec:SumOut}
Heavy ion double charge exchange reactions have been investigated with the focus on the reaction dynamics of this special class of two-step reactions. The reaction mechanism was described as a double-SCE reaction given by two consecutive SCE reaction steps which are promoted by the projectile-target residual isovector NN-interaction. The DCE reaction amplitude was constructed accordingly as a second order distorted wave matrix element. Broad space was given to disentangle nuclear matrix elements and ion-ion initial and final state interactions. As a first application of the theory, the DCE reaction $^{18}O+{}^{40}Ca\to {}^{18}Ne+{}^{40}Ar $ was investigated.

The DSCE reaction amplitude, Eq. \eqref{eq:Mab}, has the formal structure of a matrix element in second order perturbation theory, describing here the next-to-leading-order contribution of the $a+A$ residual isovector interactions. Hence, on the formal level, the DCE amplitude resembles the NME of $2\nu 2\beta$ decay. In order to understand that connection on the quantitative level, the NME of DCE reactions were studied. The central message is that DCE reactions cover at the same time dynamical aspects typical for $2\nu 2\beta$ and $0\nu 2\beta$ decay.

We emphasize that excitations of DCE modes will proceed in general by mixtures of $S=0$ non-spinflip and $S=1$ spinflip transitions, thus lifting the strict selection rules known for SCE transitions. One reason is that the final nuclear configurations in projectile and target are of $2p2h$-character with respect to the parent nuclei which allow a broad spectrum of interactions. Thus, the simplicity of SCE reactions, allowing to extract single-beta decay NME from cross section data, is not maintained in the same way for DCE reactions. Accessing DCE-NME's in full detail requires to consider additional observables which are sensitive to the spin-character of the transitions. This does not exclude exceptional, yet to be discovered cases where special configurational properties are enhancing a certain spin channel.

Significant simplifications occur when one considers specific cases, such as
$0^+_{gs} \to 0^+_{gs}$ transitions both in projectile and target. Illustrative
results have been presented in this case, together
with a comparison to available experimental data.

DCE processes are determined by many new aspects of nuclear structure and reaction dynamics which have not been under scrutiny until now. Both experiment and theory are entering into hitherto unexplored territory, posing unexpected challenges but opening a new field of nuclear research. In a forthcoming paper, a competing DCE reaction mechanism will be studied resembling the non-SSD case of Eq.\eqref{eq:MDSCE_CC} but of different dynamical origin \cite{Lenske:2019rce,Lenske:2019var}. In \cite{Santo2018} a new approach to DCE processes, based on an extended version of the IBM, was presented. An important message of the present work is that heavy ion DCE reactions are indeed the ideal tools to scrutinize nuclear DCE models under realistic conditions. Our results are encouraging 
systematic studies in this direction.

\section{Acknowledgments}
We wish to thank C. Agodi, F. Cappuzzello and M. Cavallaro for fruitful discussions.
{Partial funding of the Catania group by the European Union's Horizon 2020 research and innovation program under Grant No. 654002 is acknowledged. J.-A. L. acknowledges support by the Spanish Ministerio de Ciencia, Innovaci\'on y Universidades and FEDER funds under project FIS2017-88410-P. H.L. acknowledges gratefully partial support by Alexander-von-Humboldt Foundation, DFG, grant Le439/16, and INFN.}

\end{document}